\documentstyle[12pt]{article}

\begin{document}

\newcommand{\gsim}{ \mathop{}_{\textstyle \sim}^{\textstyle >} }
\newcommand{\lsim}{ \mathop{}_{\textstyle \sim}^{\textstyle <} }

\baselineskip 0.7cm
\abovedisplayskip 5.7mm
\belowdisplayskip 5.7mm

\renewcommand{\thefootnote}{\fnsymbol{footnote}}
\setcounter{footnote}{1}

\begin{titlepage}
\begin{flushright}
TU-467 \\
\today
\end{flushright}

\vskip 0.35cm
\begin{center}
{\Large \bf On the Solution to the Polonyi Problem with  $O$(10~TeV)
Gravitino Mass in Supergravity}
\vskip 0.5cm

\vskip 0.25cm

T.~Moroi
\footnote{Fellow of the Japan Society for the Promotion of Science.}
, M.~Yamaguchi and T.~Yanagida

\vskip 0.25cm
{\it Department of Physics, Tohoku University, Sendai 980-77, Japan}

\vskip 1.5cm

\abstract{
We study the solution to the Polonyi problem where the Polonyi field
weighs as $O(10{\rm TeV})$ and decays just before the primordial
nucleosynthesis.  It is shown that in spite of a large entropy
production by the Polonyi field decay, one can naturally explain the
present value of the baryon-to-entropy ratio, $n_B/s \sim
(10^{-10}-10^{-11})$ if the Affleck-Dine mechanism for baryogenesis
works. It is pointed out, however, that there is another cosmological
problem related to the abundance of the lightest superparticles produced
by the Polonyi decay.}

\end{center}
\end{titlepage}

\renewcommand{\thefootnote}{\arabic{footnote}}
\setcounter{footnote}{0}

$N=1$ supergravity~\cite{NPB212-413} is not only regarded as an
effective field theory of superstring below the Planck scale, but also
provides a natural framework for soft supersymmetry (SUSY)-breaking
terms at the electro-weak energy scale. Most of the supergravity models,
however, contain a light massive boson $\phi$ (Polonyi field) with the
mass $m_\phi$ of order the gravitino mass
$m_{3/2}$~\cite{PLB131-59,PRD49-779,PLB318-447}, which is responsible
for the spontaneous SUSY breaking. The Polonyi field $\phi$ couples only
gravitationally to the light particles and hence the lifetime of $\phi$
is very large as $\tau_\phi \sim M_P^2/m_\phi^3$ (with $M_P$ being the
Planck mass, $M_P = \sqrt{8\pi}M \simeq 1.2\times 10^{19}{\rm GeV}$).

This fact leads to a serious cosmological difficulty (so-called Polonyi
problem)~\cite{PLB131-59,PRD49-779,PLB318-447}.  Under quite general
assumptions, the Polonyi field $\phi$ takes an amplitude of order $M_P$
at the end of inflation, and subsequently it starts oscillation and
dominates the energy density of the universe until it decays. If the
decay of $\phi$ occurs after or during the primordial nucleosynthesis,
it most likely destroys one of the successful scenarios in the big-bang
cosmology, that is the nucleosynthesis.  Furthermore, the $\phi$ decay
releases a tremendous amount of entropy and dilutes primordial baryon
asymmetry much below what is observed today.

It has been pointed out~\cite{PLB174-176} that the first problem can be
solved by raising the Polonyi mass $m_\phi$ (or equivalently the
gravitino mass $m_{3/2}$) up to $O(10{\rm TeV})$ so that the reheating
temperature $T_R(\phi )$ by the $\phi$ decay is larger than $O(1{\rm
MeV})$. Then, the nucleosynthesis may re-start after the $\phi$ decay.
This solution favors strongly ``no-scale type''
supergravity~\cite{PRD50-2356},\footnote
{In the original no-scale supergravity
model~\cite{PLB143-410,PLB147-99}, Polonyi field acquires a mass of the
order of $m_{3/2}^2/M$~\cite{NPB241-406} which is much smaller than
the gravitino mass $m_{3/2}$. However, in the ``no-scale type''
supergravity model studied in Ref.~\cite{PRD50-2356}, the mass of the
Polonyi field is at the order of the gravitino mass.}
since the gravitino mass can be taken $O(10 {\rm TeV})$ without
diminishing the original motivation as a solution to the hierarchy
problem.  Namely, we can raise the gravitino mass while keeping all
masses of SUSY particles in observed sector to be $O(10^2{\rm GeV})$.

In this letter, we stress that the second problem can be also solved if
the Affleck-Dine mechanism~\cite{NPB249-361} for baryogenesis works in
the early universe. However, we point out another cosmological problem
that the lightest superparticles (LSPs) produced via the Polonyi decay
are so abundant that their energy density, if stable, overcloses the
universe.

The Affleck-Dine mechanism~\cite{NPB249-361} for baryogenesis is based
on the fact that there are some combinations of squark $\tilde{q}$ and
slepton $\tilde{l}$ fields for which the scalar potential vanishes
identically when SUSY is unbroken. After SUSY breaking these
flat-direction fields acquire masses $m$ of order $10^{2}$~GeV. One of
these flat directions $\chi$ is assumed to have a large initial value
$\chi_0\sim M_{GUT}$, where $M_{GUT}\sim 2\times 10^{16}{\rm GeV}$ is
the GUT scale. It has been shown~\cite{NPB249-361} that the decay of the
coherent oscillation mode of such a field $\chi$ can generate a large
baryon-to-entropy ratio $\sim O(1)$ under the presence of tiny
baryon-number violating operators such as
$(m/M_{GUT})\tilde{q}\tilde{q}\tilde{q}\tilde{l}$.

We now compute the dilution factor due to the Polonyi field $\phi$
decay. We assume, in the present analysis, that the energy density of
the universe is dominated by the coherent oscillation of $\phi$ after
the Hubble expansion rate $H$ becomes smaller than $m_\phi$.  This
assumption may be justified if the initial amplitude of $\phi$ is of the
order of the Planck scale.\footnote
{If the initial amplitude of the Affleck-Dine field $\chi$ is also
$O(M_{P})$, the energy density of $\chi$ is comparable to that of the
Polonyi field after the coherent oscillation of $\chi$ starts. Even in
this case, the following arguments are still valid.}
The Affleck-Dine field $\chi$ starts its oscillation when $H$ decreases
to its mass $m_\chi\sim O(10^2{\rm GeV})$.  The energy density $\rho$ is
given by
\begin{eqnarray}
	\rho \simeq m_\phi^2 |\phi|^2 +
	m_\chi^2 |\chi|^2 \equiv
	\kappa^2 m_\chi^2 |\chi|^2,
\end{eqnarray}
where
\begin{eqnarray}
	\kappa^2 = \frac{m_\phi^2 |\phi|^2}
	{m_\chi^2 |\chi|^2} +1.
\end{eqnarray}
Notice that $\kappa$ remains constant as far as both $\chi$ and $\phi$
obey coherent oscillation. Then $H$ is given by
\begin{eqnarray}
	H = \frac{\kappa m_\chi|\chi|}{\sqrt{3}M}.
\end{eqnarray}
When the $\chi$ starts its oscillation, the Hubble expansion rate is
about $m_\chi$, $H\simeq m_\chi$, and hence
\begin{eqnarray}
	\kappa \simeq \frac{\sqrt{3}M}{|\chi|}
	\simeq \frac{\sqrt{3}M}{|\chi_0|}.
\end{eqnarray}
Here, we have used the approximation that the amplitude of the
Affleck-Dine field $\chi$ is equal to the initial value $\chi_0$, since
the $\chi$ remains almost constant until the coherent oscillation of
$\chi$ starts.

One may expect that the decay of the field $\chi$ finishes when the
decay rate $\Gamma_\chi$ becomes comparable with the universe's
expansion rate $H$. The decay rate $\Gamma_\chi$ is estimated
as~\cite{NPB249-361}
\begin{eqnarray}
	\Gamma_\chi \sim \left( \frac{\alpha}{\pi} \right)^2
	\frac{m_\chi^3}{|\chi|^2},
\end{eqnarray}
and hence the equation $\Gamma_\chi \sim H$ gives
\begin{eqnarray}
	|\chi_D| \sim m_\chi \left( \frac{\alpha}{\pi} \right)^{2/3}
	\kappa^{-1/3} \left( \frac{M}{m_\chi} \right)^{1/3},
\end{eqnarray}
where $\chi_D$ represents the amplitude of the Affleck-Dine field at its
decay time. The reheating temperature $T_R(\chi)$ by the $\chi$ decay
would be given by
\begin{eqnarray}
	T_R(\chi) &\sim& \left( m_\chi^2 |\chi_D|^2 \right)^{1/4}
\nonumber \\ &\sim&
	m_\chi \left( \frac{\alpha}{\pi} \right)^{1/3}
	\left( \frac{\chi_0}{m_\chi} \right)^{1/6}.
\end{eqnarray}
For $\chi_0\sim M_{GUT}$ this yields $T_R(\chi)\sim 100m_\chi$.

However, as pointed out by Linde~\cite{PLB160-243}, the $\chi$ decay can
not reheat the universe up to the temperature $T\gsim m_\chi$, since the
effective masses of quarks and leptons in the thermal bath exceed
$m_\chi$ at the temperature larger than $m_\chi$. Namely, the decay
process of $\chi$ is blocked until the amplitude of $\chi$ becomes
$O(m_\chi)$. The energy density of $\chi$ at its decay time is given by
\begin{eqnarray}
	\rho_\chi (T_R(\chi )) = \frac{\pi^2}{30} g_*(T_R(\chi ))
	T_R(\chi )^4.
\end{eqnarray}
Here, $g_*(T)$ is the degree of freedom of the relativistic particles at
the temperature $T$, and we have used the energy conservation.  At this
time, the energy density of $\phi$ is given by
\begin{eqnarray}
	\rho_\phi (T_R(\chi)) =
	\rho_\chi (T_R(\chi)) \frac{\rho_\phi}{\rho_\chi} \simeq
	\kappa^2 \rho_\chi (T_R(\chi)).
\end{eqnarray}
Here we have used $\kappa\gg 1$. Furthermore, the baryon-to-entropy
ratio by the $\chi$ decay is given by
\begin{eqnarray}
	\frac{n_B}{s} =\eta_{B0} \frac{n_\chi}{s} =
	\frac{3}{4} \frac{T_R(\chi )}{m_\chi} \eta_{B0}.
\end{eqnarray}
where $\eta_{B0}\equiv n_B/n_\chi$, $s$ is the entropy density of the
thermal bath
\begin{eqnarray}
	s(T) \equiv \frac{2\pi^2}{45} g_*(T) T^3,
\end{eqnarray}
and $n_\chi$ is the number density of $\chi$ which is given by $n_\chi
\equiv \rho_\chi/m_\chi$. With $T_R(\chi )\sim m_\chi$, $n_B/s \sim
\frac{3}{4} \eta_{B0}$, which is expected to be of order unity provided
$\eta_{B0}\sim O(1)$.

The dilution factor $D$ due to the Polonyi field $\phi$ decay can be
easily calculated. From the energy conservation, the reheating
temperature $T_R(\phi )$ by the $\phi$ decay is given by
\begin{eqnarray}
	T_R(\phi )^4 = \frac{90}{\pi^2g_*(T_R(\phi ))}
	\Gamma_\phi^2 M^2,
\end{eqnarray}
where $\Gamma_\phi \sim Nm_\phi^3/M_P^2$ is the decay width of the
Polonyi field with $N\sim 100$ being the number of the decay channel.
Here we have used a naive approximation that the Polonyi field $\phi$
decays instantaneously at $H=\Gamma_\phi$.\footnote
{In fact, the decay of the Polonyi field does not occur instantaneously.
However as pointed out in Ref.~\cite{PRD31-681}, results of our naive
approximation are not so different from the exact ones.}
Then the reheating temperature $T_R(\phi)$ is estimated to be $O(1{\rm
MeV})$ for $m_\phi =O(10{\rm TeV})$. The temperature of the thermal bath
$T_f$ before the $\phi$ decay is given by
\begin{eqnarray}
	\frac{g_*(T_f)}{g_*(T_R(\chi ))}
	\left( \frac{T_f}{T_R(\chi )} \right)^3 =
	\left( \frac{R(T_R(\chi ))}{R(T_R(\phi ))} \right)^3 =
	\left( \frac{\rho_\phi (T_f)}
	{\rho_\phi (T_R(\chi))} \right),
\end{eqnarray}
where $R$ is the scale factor of the universe, and $\rho_\phi (T_f) =
3\Gamma_\phi^2 M^2$. Then we obtain the dilution factor $D$
\begin{eqnarray}
	D &=& \frac{g_*(T_f)}{g_*(T_R(\phi ))}
	\left( \frac{T_f}{T_R(\phi )} \right)^3
\nonumber \\ &=&
	\frac{g_*(T_f)}{g_*(T_R(\phi ))}
	\frac{\rho_\phi (T_f)}{\rho_\phi (T_R(\chi))}
	\left( \frac{T_R(\chi )}{T_R(\phi )} \right)^3
\nonumber \\ &=&
	\frac{T_R(\phi )}{T_R(\chi )}
	\frac{\chi_0^2}{3M^2}
\nonumber \\ &\sim&
	10^{-5} \left( \frac{\chi_0}{M} \right)^2,
\label{dilution}
\end{eqnarray}
for $T_R(\phi ) \sim O(1{\rm MeV})$. We finally obtain a net baryon
asymmetry in the present universe
\begin{eqnarray}
	\frac{n_B}{s} = \frac{3}{4} \eta_{B0} D \sim
	10^{-5} \eta_{B0} \frac{\chi_0^2}{M^2},
\label{eta_now}
\end{eqnarray}
with which one may explain the observed value $n_B/s \sim (10^{-10} -
10^{-11})$ taking $\chi_0 \sim M_{GUT}$, $T_R(\phi )\sim 1{\rm MeV}$,
and $\eta_{B0}\sim 1$.

We should comment that the final result (\ref{eta_now}) can be more
easily derived if one notices that $n_B/\rho_\phi$ is independent of
time since the baryon number is approximately conserved in the regime we
consider. Then,
\begin{eqnarray}
	\frac{m_\chi n_B}{\rho_\phi} = {\rm const}.
\label{constant}
\end{eqnarray}
We evaluate this when the Affleck-Dine field $\chi$ starts its
oscillation. At this time,
\begin{eqnarray}
	\frac{m_\chi n_B}{\rho_\phi} = \frac{m_\chi n_B}{\rho_\chi}
	\frac{\rho_\chi}{\rho_\phi}
	=
	\eta_{B0} \frac{\rho_\chi}{\rho_\phi}
	=
	\eta_{B0} \left( \frac{\chi_0}{\sqrt{3}M} \right)^2.
\label{at_oscillation}
\end{eqnarray}
Here we have used $H\simeq \sqrt{\rho_\phi}/\sqrt{3}M$, and $\rho_\chi =
m_\chi^2\chi_0^2$. On the other hand, we evaluate the same quantity
given in Eq.~(\ref{constant}) at the decay time of the Polonyi field
$\phi$
\begin{eqnarray}
	\frac{m_\chi n_B}{\rho_\phi} = \frac{4}{3}
	\frac{m_\chi n_B(T_R(\phi ))}{s(T_R(\phi )) T_R(\phi )}.
\label{at_decay}
\end{eqnarray}
Equating Eq.~(\ref{at_oscillation}) and Eq.~(\ref{at_decay}), we get
Eq.~(\ref{eta_now}).

The dilution factor $D$ we have obtained is much larger than that
derived in the previous work~\cite{PLB174-176}. For example, the
dilution factor given in Ref.~\cite{PLB174-176} is $\sim 10^{-14}$ for
the case $T_R(\phi ) \sim 1{\rm MeV}$, which is about $10^{-9}$ times
smaller than our result with $T_R(\chi ) \sim m_\chi \sim 100{\rm GeV}$
and $\chi_0 \sim M$. This discrepancy originates to the fact that the
amplitude of the Polonyi field has already decreased by a large amount
at the decay time of the Affleck-Dine field. In fact, we find that when
the Affleck-Dine baryogenesis occurs ({\it i.e.} $T=T_R(\chi)\simeq
m_\chi$), the amplitude of the Polonyi field $\phi$ is $\phi(T_R(\chi ))
\sim (m_\chi^2 M/m_\phi\chi_0)\ll M$, which gives
$\rho_\phi(T_R(\chi )) \sim (m_\chi^2M/\chi_0)^2$. One can reconstruct
the final form of Eq.~(\ref{dilution}) by using this effective initial
amplitude $\phi(T_R(\chi))$. In Ref.~\cite{PLB174-176}, this effect is
not taken into account, and hence the dilution factor given in
Ref.~\cite{PLB174-176} is underestimated.

Let us now turn to discuss a new cosmological difficulty in the present
solution to the Polonyi problem. Through the decay of the Polonyi field, an
amount of superparticles are produced, which promptly decay to the
lightest superparticle (LSP).  The number density of the LSPs,
$n_{LSP}$, produced by these processes is roughly the same as that of
the Polonyi fields.  The ratio of $n_{LSP}$ to the entropy density is
thus found to be
\begin{equation}
	\left. \frac{n_{LSP}}{s} \right|_{T=T_R(\phi )} =
	\frac{n_\phi}{s}=\frac{\rho_\phi}{m_\phi s}
	\simeq \frac{T_R}{m_\phi}.
\label{nLSP/s}
\end{equation}
Under the assumption of $R$-parity conservation, only pair annihilation
processes can change the number of the LSPs in the comoving volume.  If
it does not take place, the ratio (\ref{nLSP/s}) remains constant, and
so does the ratio of their energy density to the entropy density.  From
this it follows that
\begin{equation}
	\left.\frac{\rho_{LSP}}{s}\right|_{\rm Today}\simeq
	m_{LSP} \frac{T_R(\phi )}{m_\phi}
	 = 10^{-5} {\rm GeV} \left(\frac{m_{LSP}}{100 {\rm GeV}} \right)
        \left(\frac{T_R(\phi )}{1 {\rm MeV}} \right)
  	 \left( \frac{10^4 {\rm GeV}}{m_\phi} \right),
\label{rhoLSP/s}
\end{equation}
where $m_{LSP}$ is the mass of the LSP. This value should be compared
with the ratio of the critical density, $\rho_{cr.}$, to the entropy
density, $s_0$, in the present universe
\begin{equation}
\frac{\rho_{cr.}}{s_0}=3.6 h^2 \times 10^{-9} {\rm GeV},
\label{rhocr/s}
\end{equation}
with $h$ $(0.5 \leq h \leq 1)$ being the Hubble constant in units of
$100$ km/sec/Mpc.  We can see that the relic abundance of the LSPs is
several orders of magnitude larger than the critical density, unless the
mass of the LSP is not very light $m_{LSP} \lsim 10$ MeV. In the minimal
supersymmetric standard model (MSSM), the LSP is most likely in the
neutralino sector which consists of a bino, a neutral wino and two
neutral higgsinos.  The negative searches for the neutralinos at LEP
give constraints on the neutralino sector.  Assuming the GUT relation of
the gaugino masses, we obtain the lower limit of the neutralino mass of
about 18 GeV~\cite{PRD44-927}.  Thus, we need to discuss whether
annihilation processes can reduce their relic density not to exceed the
closure limit (\ref{rhocr/s}).

When the annihilation of the LSPs occurs, its number density can be
reduced to
\begin{equation}
	n_{LSP} \simeq
	\left. \frac{H}{\langle \sigma_{ann} v_{rel} \rangle}
	\right|_{T=T_R(\phi )},
\end{equation}
where $\langle \sigma_{ann} v_{rel} \rangle$ is the thermal average of
the annihilation cross section $\sigma_{ann}$ times the relative
velocity $v_{rel}$.  In this case, the relic abundance of the LSPs at
the present day is calculated to be
\begin{equation}
	\left. \frac{\rho_{LSP}}{s} \right|_{\rm Today} \simeq
	g^{-1/2}_* (T_R(\phi )) \frac{m_{LSP}}{T_R(\phi )}
	\frac{1}{\langle \sigma_{ann} v_{rel} \rangle M}.
\end{equation}
Comparing this equation with Eq.~(\ref{rhocr/s}), we find that the
annihilation cross section must satisfy
\begin{equation}
	\langle \sigma_{ann} v_{rel} \rangle \gsim
	3 h^{-2} \times 10^{-7} {\rm GeV^{-2}}
	\left( \frac{m_{LSP}}{10^2 {\rm GeV}} \right)
	\left( \frac{10 {\rm MeV}}{T_R(\phi )} \right),
\label{required-cross-section}
\end{equation}
in order that the energy density of the LSPs should not overclose the
Universe.

We now examine if the cosmological constraint
(\ref{required-cross-section}) is satisfied in the MSSM. When the LSP is
bino-dominant, it annihilates to a lepton pair via the exchange of
right-handed sleptons.  The cross section is $p$-wave dominant, giving
\cite{PLB230-78}
\begin{equation}
	\langle \sigma_{ann} v_{rel} \rangle = 2 \pi \alpha'^2\left\{
	\frac{4m_{LSP}^2}{(m_{LSP}^2+m_{\tilde l_{R}}^2)^2}
	- \frac{8m_{LSP}^4}{(m_{LSP}^2+m_{\tilde l_{R}}^2)^3}
	+ \frac{8m_{LSP}^6}{(m_{LSP}^2+m_{\tilde l_{R}}^2)^4}
	\right\}  \langle v^2 \rangle.
\label{bann1}
\end{equation}
Here $\alpha' =g^{\prime 2} / 4 \pi \simeq 0.01$ is the fine structure
constant for the $U(1)_Y$ gauge interaction, $m_{\tilde l_R}$ the mass
of the right-handed slepton and $\langle v^2 \rangle$ represents the
average of the squared velocity of the LSP, which could be determined by
a close inspection of the scattering of the LSP off the thermal
radiations.  For the case of $E\gg m_{LSP}$, we estimate the elastic
scattering cross section off an electron in the thermal bath to be of
the order of
\begin{equation}
	\langle \sigma v_{rel} \rangle \sim  \alpha^{\prime 2}
	\frac{E T_R(\phi )}{m_{\tilde e_R}^4},
\end{equation}
where $E$ is the energy of the LSP,  yielding  the interaction rate
$\Gamma$ versus the expansion rate $H$
\begin{equation}
	\frac{\Gamma}{H} \sim \alpha^{\prime 2}
	\frac{E T_R(\phi )^2 M}{m_{\tilde e_R}^4} \sim
	10^4 \left( \frac{E}{10^2 {\rm GeV}} \right).
\end{equation}
This implies that by a series of scattering, the LSPs, which were
energetic when produced, lose their momenta substantially and they will
go to the non-relativistic energy region.  Once entering in the
non-relativistic region, however, their momentum loss is at most about
the background temperature at one collision and they do not likely reach
their kinetic equilibrium.  Therefore we expect\footnote
{If the LSP is in kinetic equilibrium, the average $\langle v^2\rangle$
is of order $T_R(\phi )/m_{LSP} \sim 10^{-4}$.}
\begin{equation}
    \frac{T_R(\phi )}{m_{LSP}} \ll \langle v^2 \rangle \ll 1.
\end{equation}
Even with our ignorance of the precise value of $\langle v^2 \rangle$,
we can see that the annihilation cross section is not large enough.
Indeed, taking $m_{\tilde l_R} = m_{LSP} \sim 10^2$ GeV,
Eq.~(\ref{bann1}) leads
\begin{equation}
	\langle \sigma_{ann} v_{rel} \rangle =
	\pi \alpha^{\prime 2} \frac{1}{m_{LSP}^2} \langle v^2\rangle
	\sim 3  \langle v^2 \rangle \times 10^{-8} {\rm GeV^{-2}},
\end{equation}
which is smaller than the requisite annihilation cross section
(\ref{required-cross-section}) even for $\langle v^2 \rangle \sim
1$.\footnote
{We can apply a similar argument to a photino-like LSP.}
Notice that if the LSP can annihilate into a top-quark pair in the
$s$-wave, the cross section becomes
\begin{eqnarray}
	\langle \sigma_{ann} v_{rel} \rangle =
	\frac{32 \pi}{27} \alpha^{\prime 2}
	\frac{m_t^2}{(m_{\tilde t_R}^2+m_{LSP}^2-m_t^2)^2}
	\left( 1-\frac{m_t^2}{m_{LSP}^2} \right)^{1/2},
\end{eqnarray}
where $m_t$ and $m_{\tilde t_R}$ are the masses of top-quark and
right-handed stop, respectively.  This cross section takes its maximum
about $3 \times 10^{-9}$ GeV$^{-2}$, which is again not large enough to
reduce the number density of the LSPs sufficiently.

Next consider the case where the LSP is higgsino-like.  Since the
interaction of the (neutral) higgsino with particles in the thermal
bath, {\it i.e.} a photon and an election, is very weak, the energy loss
of the higgsino is not effective and it will remain in the relativistic
region.  This higgsino dominantly annihilate into a $W$-boson pair whose
cross section depends on the energy in the center of mass frame
$\sqrt{s}$.  Though we do not know the precise value of $\sqrt{s}$, the
upperbound on the cross section is given by taking $\sqrt{s} =2m_\chi$
\cite{PLB230-78}, that is
\begin{equation}
	\langle \sigma_{ann} v_{rel} \rangle \lsim
	\frac{\pi\alpha_2^2}{2} \frac{m_\chi^2}{(2m_\chi^2-m_W^2)^2}
        \left( 1-\frac{m_W^2}{m_\chi^2} \right)^{3/2},
\label{higgsino-cross-section}
\end{equation}
where $\alpha_2$ is the fine structure constant for the $SU(2)_L$ gauge
interaction. This takes the maximum $\sim 2 \times 10^{-8} $ GeV$^{-2}$
and again it does not satisfy the requirement
(\ref{required-cross-section}).  Similar arguments apply to the case
where the LSP is an admixture of the gaugino and the higgsino.  Recall
that the annihilation cross section in the mixed region is generically
smaller than that in the higgsino case (\ref{higgsino-cross-section})
\cite{PRD41-3565}.  One may think that the annihilation via the Higgs
exchange \cite{PLB245-545} can enhance its cross section if $s$ hits the
$s$-channel pole.  However, since the LSPs are not in the kinetic
equilibrium, their momentum distribution is rather wide and hence the
annihilation through this $s$-channel pole will not be effective enough
to reduce the relic density.  Summarizing these arguments, we conclude
that, in the MSSM with the $R$-parity conservation, the relic abundance
of the LSPs will always be too large to be cosmologically viable. The
relic density can be reduced to a cosmologically acceptable level, if
one raises the reheating temperature $T_R(\phi )$ to $O(10$
GeV).\footnote
{For the case $10{\rm MeV} \lsim T_R(\phi ) \lsim 10{\rm GeV}$, the
constraint (\ref{required-cross-section}) becomes weaker. In this case
one needs to know $\langle v^2\rangle$ precisely in order to know the
relic density of the LSP. This requires a detailed analysis on the time
evolution of the distribution function of the LSP. This is beyond the
scope of this letter, and will be given elsewhere.}
However, it means that the gravitino is heavier than $O(10^6$ GeV).
Such a large gravitino mass will be, perhaps, disfavored from the
naturalness point of view.

To cure this conflict in the case of $T_R(\phi ) \lsim 10$ MeV, let us
consider modifications of the MSSM.  One way is to extend the particle
contents and provide a new, very light LSP.  If the LSP is lighter than
$O(10$ MeV), we can see from Eq.~(\ref{rhoLSP/s}) that the relic
abundance does not exceed the critical density without invoking the
annihilation.  This is most easily realized in the minimal extension of
the MSSM, where the superpartner of a singlet Higgs is contained in the
neutralino sector.  A recent analysis shows that even a massless LSP is
still an allowed possibility in this model \cite{FFB}.  Another
extension which has a light LSP is to incorporate the Peccei-Quinn
symmetry.  Then the superpartner of the axion, the axino, can be the
LSP~\cite{NPB358-447}.  Indeed, it was shown in Ref.~\cite{PLB276-103}
that the axino becomes massless at the tree-level in the no-scale
supergravity.  Radiative corrections may give a small, model-dependent
axino mass.\footnote
{A light axino can also be realized if one chooses a special form of
superpotential~\cite{PLB287-123}.}
In the case of the axino mass $\sim 10$ MeV, the axino
becomes a cold dark matter of the universe.

$R$-parity breaking is the other possibility to make our scenario
cosmologically viable. In this case, the LSP is no longer stable, but
decays to ordinary particles.  If the lifetime $\tau_{LSP}$ of the LSP
is shorter than 1~sec,\footnote
{Such a small $R$-parity violation ($\tau_{LSP} \sim 1{\rm sec}$) is
consistent with other phenomenological constraints~\cite{PLB256-457}.}
its decay does not upset the standard big-bang nucleosynthesis.


\newpage
\begin{thebibliography}{99}
%
\bibitem{NPB212-413}
	E.~Cremmer, S.~Ferrara, L.~Grardello and A.~van~Proeyen,
	{\sl Nucl.Phys.} {\bf B212} (1983), 413.
%
\bibitem{PLB131-59}
	 G.D.~Coughlan, N.~Fischler, E.W.~Kolb, S.~Raby and G.G.~Ross,
	{\sl Phys.Lett.} {\bf 131B} (1983) 59.
%
\bibitem{PRD49-779}
	T.~Banks, D.B.~Kaplan and A.E.~Nelson,
	{\sl Phys.Rev.} {\bf D49} (1994) 779.
%
\bibitem{PLB318-447}
	B.~de~Carlos, J.A.~Carlos, F. Quevedo and E.~Roulet,
	{\sl Phys.Lett.} {\bf B318} (1993) 447.
%
\bibitem{PLB174-176}
        J.~Ellis, D.V.~Nanopoulos and M.~Quiros,
        {\sl Phys.Lett.} {\bf B174} (1986) 176.
%
\bibitem{PRD50-2356}
	H.~Murayama, H.~Suzuki, T.~Yanagida and J.~Yokoyama,
	{\sl Phys.Rev.} {\bf D50} (1994) 2356.
%
\bibitem{PLB143-410}
	J.~Ellis, C.~Kounnas and D.V.~Nanopoulos,
	{\sl Phys.Lett.} {\bf B143} (1984) 410.
%
\bibitem{PLB147-99}
	J.~Ellis, K.~Enqvist and D.V.~Nanopoulos,
	{\sl Phys.Lett.} {\bf B147} (1984) 99.
%
\bibitem{NPB241-406}
	J.~Ellis, C.~Kounnas and D.V.~Nanopoulos,
	{\sl Nucl.Phys.} {\bf B241} (1984) 406.
%
\bibitem{NPB249-361}
	I.~Affleck and M.~Dine,
	{\sl Nucl.Phys.} {\bf B249} (1985), 361.
%
\bibitem{PLB160-243}
	A.D.~Linde,
	{\sl Phys.Lett.} {\bf B160} (1985) 243.
%
\bibitem{PRD31-681}
	R.J.~Scherrer and M.S.~Turner,
	{\sl Phys.Rev.} {\bf D31} (1985) 681.
%
\bibitem{PRD44-927}
        K. Hidaka,
	{\sl Phys. Rev.}{\bf D44}, 927 (1991).
%
\bibitem{PLB230-78}
	K.A.~Olive and M.~Srednicki,
	{\sl Phys.Lett.} {\bf B230} (1989) 78.
%
\bibitem{PRD41-3565}
	K.~Griest,  M.S.~Turner and M.~Kamionkowski,
	{\sl Phys.Rev.} {\bf D41} (1990) 3565.
%
\bibitem{PLB245-545}
	J.~Ellis, L.~Roskowski and Z.~Lalak,
	{\sl Phys.Lett.} {\bf B245} (1990) 545.
%
\bibitem{FFB}
	F.~Franke, H.~Fraas and A.~Bartl,
	preprint UWITP-94/2, UWThPh-1994-17 (July 1994).
%
\bibitem{NPB358-447}
	K.~Rajagopal, M.S.~Turner and F.~Wilczek,
	{\sl Nucl.Phys.} {\bf B358} (1991) 447.
%
\bibitem{PLB276-103}
	T.~Goto and M.~Yamaguchi,
	{\sl Phys.Lett.} {\bf B276} (1992) 103.
%
\bibitem{PLB287-123}
	E.J.~Chun, J.E.~Kim and H.P.~Nilles,
	{\sl Phys.Lett.} {\bf B287} (1992) 123.
%
\bibitem{PLB256-457}
        B.A.~Campbell, S.~Davidson,~J.~Ellis and K.A.~Olive,
        {\sl Phys.Lett.} {\bf B256} (1991) 457.
%
\end{thebibliography}
\end{document}